\documentstyle[11pt,aaspp4]{article}
\begin{document}
\title{Possible Evidence for Relativistic Shocks in Gamma-Ray Bursts}
\notetoeditor{Until June 1, 1997 address correspondence to J. I. Katz c/o
T. Piran, Racah Institute of Physics, Hebrew University, Jerusalem 91904,
Israel.  After June 1, 1997 address correspondence to J. I. Katz,
Department of Physics, Washington University, St. Louis, Mo. 63130.}
\author{E. Cohen, J. I. Katz\altaffilmark{1}, T. Piran, R. Sari}
\affil{Racah Institute of Physics, Hebrew University, Jerusalem 91904, 
Israel}
\authoremail{udic@cc.huji.ac.il}
\authoremail{katz@wuphys.wustl.edu}
\authoremail{tsvi@shemesh.fiz.huji.ac.il}
\authoremail{sari@shemesh.fiz.huji.ac.il}
\author{R. D. Preece}
\affil{Department of Physics, University of Alabama at Huntsville, 
Huntsville, Ala. 35899}
\authoremail{rob.preece@msfc.nasa.gov}
\and
\author{D. L. Band}
\affil{Center for Astrophysics and Space Sciences, UCSD, La Jolla, Cal. 
92093}
\authoremail{dlbbat@cass09.ucsd.edu}
\altaffiltext{1}{Permanent address: Department of Physics and McDonald 
Center for the Space Sciences, Washington University, St. Louis, Mo. 63130}
\begin{abstract}
Relativistic shock models of gamma-ray bursts may be tested by comparison of
their predicted low energy asymptotic spectral indices $s$ to observations.
Synchrotron radiation theory predicts that the instantaneous spectrum has 
$s = - 1/3$ and the spectrum integrated over the radiative decay of the 
electrons' energies has $s = 1/2$, with other cases lying between these 
limits.  We examine the spectra of 11 bursts obtained by the Large Area 
Detectors on BATSE.  One agrees with the predicted instantaneous spectrum, 
as does the initial portion of a second, and three are close to the 
predicted integrated spectrum.  All the observed asymptotic spectral slopes 
lie in the predicted range.  This evidence for relativistic shocks is 
independent of detailed models of bursts and of assumptions about their 
distances.  Radiation observed with the predicted instantaneous spectrum
has a comparatively smooth time dependence, consistent with the necessarily
long radiation time, while that with the predicted integrated spectrum has a
spiky time dependence, consistent with the necessarily short radiation time.
\end{abstract}
\section{Introduction}
The high intensities, likely large distances, short time-scales and inferred
small source sizes of gamma-ray bursts (GRB) have led most astrophysicists
to conclude that they involve relativistic motion, and that their radiation
is produced when the kinetic energy of this motion is thermalized in a
relativistic shock.  \cite{K94a} and \cite{T95} argued that acceleration by 
a relativistic shock produces a characteristic distribution of particle 
energies.  Because all the particles are accelerated, a simple phase space 
argument implies that the energy distribution function (in the frame of the 
shocked fluid) is peaked about the mean internal energy per particle (in 
that frame).  This characteristic energy is defined by the ratio of the 
proper energy density to the proper number density.  Thus, nearly all the 
particles are relativistic.  Relativistic shock acceleration is 
characterized by a paucity of low energy particles and by a synchrotron 
spectrum which, below the characteristic synchrotron frequency (of a 
particle with the characteristic energy), resembles that of a monoenergetic 
distribution of particles.  Therefore, the instantaneous synchrotron 
spectrum of particles accelerated by a relativistic shock should always (in 
the absence of absorption) have a low frequency asymptotic spectral index 
$s = - \onethird$ ($F_\nu \propto \nu^{-s} = \nu^{1/3}$).  This result 
explains the well known ``X-ray paucity'' problem of GRB, and is independent
of the presence of a high energy ``tail'' to the particle distribution 
function.

These predictions may be compared to those of stochastic acceleration in
non-relativistic shocks, in which no such characteristic energy can be
defined because there is an effectively infinite reservoir of unaccelerated
thermal particles.  The proper number density of accelerated particles is
indeterminate, because additional thermal particles can be recruited to
the accelerated distribution (if the total number density, including
thermal particles, were used to define the characteristic energy, the result
would be nonrelativistic---we use this definition of a non-relativistic
shock---and would tell nothing about the distribution function of the
relativistic particles).  In the absence of a characteristic energy a
power-law (rather than a peaked) particle distribution must be obtained,
extending down to a very low thermal, rest mass or injection energy.   
Radiation by these intermediate energy particles dominates the lower 
frequency part of the synchrotron spectrum of a non-relativistic shock, and 
gives it a spectral index which depends on the distribution function of 
accelerated particles.

Estimates of the synchrotron radiation efficiency and electron lifetime,
assuming magnetic equipartition (\cite{K94b,SNP96,SP97a}) in GRB suggest 
that the radiating electrons lose most of their energy by synchrotron 
radiation in a time much shorter than the duration of a typical burst, and 
especially of long bursts (the data reported in this paper all come from 
long bursts, with $T_{90} > 20$ s).  Therefore, for comparison with spectra 
integrated over the duration of a GRB the theoretical spectrum should be 
integrated over the spectrum emitted by a single electron as its energy 
decays.  Noting that the characteristic synchrotron frequency varies 
$\propto E^2$, where $E$ is the electron's energy, we have $F_\nu \propto 
dE/d\nu \propto \nu^{-1/2}$; the predicted spectral index $s = \onehalf$.  
This result is also applicable to Compton scattering if the photon field 
upon which the electrons scatter does not change as the electrons lose their 
energy.

In general, a real astronomical object (probably heterogeneous in both space
and time) will have a spectrum somewhere between these two extremes.  
Relativistic shock theory thus predicts a spectral index
$$- {1 \over 3} \le s \le {1 \over 2} \eqno(1)$$  
below the characteristic synchrotron frequency.

In a given magnetic field, an electron's radiative loss time increases
$\propto E^{-1}$ as $E \to 0$, so that at sufficiently low frequency any
observation of specified duration becomes effectively instantaneous, with a
predicted asymptotic spectral index
$$s = - {1 \over 3}. \eqno(2)$$
It is only necessary that the duration of emission be limited by radiative
energy losses and that the length of integration not increase along with the
(expected) increase of duration with decreasing frequency.  For comparison, 
non-relativistic shocks can lead to any $s \ge - \onethird$ (this lower 
bound is set by the synchrotron spectrum of monoenergetic electrons).

The purpose of this paper is to compare the predictions of relativistic
shock theory to observed spectra of GRB.  Characteristic synchrotron
frequencies (the peaks of $\nu F_\nu$ or breaks in a $\log$-$\log$ plot
of $F_\nu$) are typically in the range 300--1000 KeV.  The behavior of GRB
spectra at yet higher energies reflects the high energy ``tail'' of the
particle distribution function, and is not relevant here.  We instead
concentrate on observations at lower energies, and ask the question: Do the 
observed spectra asymptotically approach the predicted power laws of 
spectral indices $s = - \onethird$ or $s = \onehalf$?  We compare these 
predictions to spectral data obtained by BATSE's Large Area Detectors on 
CGRO.  To the extent that the predictions are confirmed, this may be the 
first empirical evidence for relativistic shocks in astrophysics.

In \S2 we report and describe data from 11 GRB.  \S3 contains a comparison
of these data to the asymptotic power laws described above.  In \S4 we show
that, within our sample, bursts with $s \approx -\onethird$ have smoother
time profiles than those with $s \approx \onehalf$, in agreement with 
expectations about their radiation times.  \S5 contains a brief
summary, and discusses relativistic shocks in other astronomical objects,
such as pulsar winds, AGN and double radio sources.
\section{Spectra of 11 GRB}
The BATSE LAD is a NaI(Tl) scintillator, whose intrinsic energy resolution
for soft gamma-rays is not high.  The fractional dispersion in pulse height 
of such scintillators for 100 KeV gamma-rays is about 10\% (\cite{OK61}).
It is larger at lower photon energies, although not by as much as a naive 
$\nu^{-1/2}$ dependence would imply.  The LAD has a FWHM (2.35 times the
dispersion for a Gaussian profile) of 27\% at 88 KeV (\cite{H91}), close to
O'Kelley's estimate.  Attempts to deconvolve the measured count distribution
to obtain the source photon spectrum by multiplying the count vector by the 
inverse of the known Detector Response Matrix (DRM) (\cite{P95}) fail 
because they amount to attempts to recover information lost in convolving 
the spectrum with the DRM.  

The usual procedure in spectral studies of GRB (\cite{S94,B96}), which we
adopt, is first to fit a standard spectral form to the data.  We use the
four parameter model described by \cite{B93}.  This model consists of two
asymptotically power law segments smoothly joined and parameterized so that
the energy of the peak in $\nu F_\nu$ characterizes the break between the
two.  It can accommodate a considerable amount of curvature in the spectrum,
and has been successfully fit to a large number of GRB.  The forward-folding
fitting and calibration procedures used here are described elsewhere 
(\cite{P97}) and incorporate the full detector response, as well as effects 
of scattering of gamma-rays in the spacecraft and the Earth's atmosphere 
(\cite{P95}).

This fitted spectral form is then folded through the DRM to obtain a predicted 
distribution of counts in each energy channel.  The ratio of the model 
photon flux to the predicted counts is then an estimate of the reciprocal
detector efficiency in each channel.  The observed
count rate in each channel is then multiplied by this reciprocal efficiency
to produce an estimate of the incident photon spectrum, which we plot.  

This procedure would be exact (independent of any assumed spectral model) if
the DRM were diagonal, and provides an estimate of the diagonal elements of 
the inverse matrix.  It is stable.   It accounts for the gross variation of 
detector efficiency with photon energy without performing an unstable 
deconvolution.  However, it is not a unique resolution of the ambiguity 
introduced by a detector response function of finite width; it cannot 
reverse the spectral broadening.  There can be no unique reversal because 
information has been irretrievably lost.  Unfortunately, the necessary 
assumption of a spectral model introduces an irremediable bias into the 
estimated reciprocal efficiencies, and this and the implicit approximation 
of the DRM as diagonal introduce bias into the estimated incident photon 
spectrum.

Figures 1a--l show the spectra of 11 GRB observed by the Large Area
Detectors on BATSE.  In general, most of the fluence was integrated to 
obtain these spectra.  They are shown in chronological order.  These bursts 
were chosen on the basis of their total fluence ranking, as determined in 
the BATSE 3B Catalog (\cite{M96}).  The calculation of fluence assumes 
that certain BATSE data types exist for each burst which cover the entire 
event; this is true for most of the bursts in the Catalog.  For two of 
the brighter bursts, the total spectrum was used for calibration of the 
Large Area Detectors and thus cannot be used for any other purpose; these 
were omitted from our list.
\placefigure{Fig1}

In the case of GRB910601 two spectra are shown.  The spectrum denoted 
GRB910601rise was obtained from the initial rising part of the burst, not
including its main peak (see Figure 2a).  This was initially done 
inadvertently, but turns out to be illuminating, as discussed in \S4.

In each figure the solid line corresponds to $F_\nu \propto \nu^{1/3}$ and
the dashed line to $F_\nu \propto \nu^{-1/2}$.  These straight lines have
been, somewhat arbitrarily, normalized to the geometric mean of the four
lowest energy data points; of course, it is their slope, and not their
normalization, which is of interest.  The horizontal error bars show the
nominal widths of the spectral bins, with the data points plotted at the
geometric mean energies.  The vertical error bars represent 1-$\sigma$
counting statistics only, and do not include any systematic errors, 
including the uncertainty introduced by the assumption of a specific 
functional form in the fits.  The absolute sensitivity calibration is 
irrelevant to testing our asymptotic models, but differential calibration
errors between spectral bins would be important; the calibration procedures
were described by \cite{P95}.  The data shown represent integrations over
most of the gamma-ray fluences of the bursts.
\section{Testing the relativistic shock model}
We are testing the predictions of the low energy {\it asymptotic} behavior
of the GRB spectrum.  Because we cannot specify the complete spectrum
(we make no predictions as to the shape of the electron distribution 
around its peak) we cannot predict at what energy, or with what shape, the
spectrum approaches the asymptotic power law.  As a result, the test of the
predicted asymptotic behavior is qualitative: do the spectra, to the eye,
appear to approach the predicted asymptotic power laws?

GRB910601rise and GRB921123 closely approach the $F_\nu \propto \nu^{1/3}$ 
asymptote at low energies.  An additional eight spectra, GRB910503, 
GRB910601, GRB920406, GRB920622, GRB930201, GRB930916, GRB940206 and 
GRB940302, may be approaching that asymptote, but the evidence is suggestive
rather than persuasive.  The remaining two bursts, GRB930506 and GRB940217, 
show no evidence of approaching this asymptote, but even these data are 
consistent with the prediction that $s \ge - \onethird$.

The instantaneous radiation of a relativistic shock was predicted to have an
asymptotic low energy spectral index of $- \onethird$.  However, the 
predicted synchrotron energy loss times are much shorter than the burst 
durations and integration times.  The confirmation of the predicted 
instantaneous asymptotic spectrum in GRB910601rise and GRB921123 suggests 
that the synchrotron loss times may have been underestimated (for example, 
if the magnetic field is below its equipartition value), or that the 
radiation is terminated (at the price of reduced radiative efficiency) by 
some other process, such as adiabatic expansion of a dense clump of 
radiating particles and field, before the particles' energy is degraded by 
radiative energy loss and their emitted spectrum is softened.

The effects of adiabatic energy loss are easy to estimate.  For
$d$-dimensional expansion ($d = 1$ corresponds to an expanding sheet or
slab, $d = 2$ to an expanding line or filament and $d = 3$ to expansion
from a point in all directions), assuming the particle distribution function
remains isotropic, a particle's energy scales with expansion distance $r$ 
and time $t$ as $E \propto r^{-d/3} \propto t^{-d/3}$.  We assume that the 
magnetic field also remains statistically isotropic, so that its energy 
density adiabatically declines like that of a relativistic fluid, and $B 
\propto r^{-2d/3} \propto t^{-2d/3}$.  If, instead, a flux conservation 
argument were used then for $d = 1$ or $d = 2$ the field would become 
strongly anisotropic and the total magnetic energy would diverge; this is 
impossible, and magnetic reconnection probably limits the magnetic energy 
and isotropizes the field.

Given the assumptions of isotropy, integration of the radiation emitted by
a particle as its energy and the magnetic field undergo adiabatic decay, 
until radiation is cut off when the particle's characteristic synchrotron 
frequency equals the frequency of observation, leads to a spectral index 
$$s = {3 - 2d \over 4d}. \eqno(3)$$
For $d = 3$ $s = -\onequarter$, essentially indistinguishable from the 
instantaneous spectrum; for
$d = 2$ $s = - 1/8$; for $d = 1$ $s = \onequarter$, not far from the
spectrum integrated over radiative energy losses.  All these spectral
indices lie between those obtained for the instantaneous spectrum and for
the spectrum integrated over the radiative decay of the electrons' energy.

GRB930201 closely approaches an asymptotic spectral index of $\onehalf$ 
(except perhaps for its lowest energy point). This may be evidence for the 
observation of radiation from a relativistic shock in which the radiating 
electrons are observed throughout their loss of energy (by radiation rather 
than by adiabatic expansion), as their characteristic synchrotron frequency 
passes through the band of observation.  GRB940302 similarly, but not so
closely, approaches an asymptotic spectral index of $\onehalf$ (again, 
except perhaps for its lowest energy point), and GRB940217 may also show 
this behavior.  

The deviations of the lowest energy points, if real, may be explained by a 
gradual transition to an asymptotic spectral index of $-\onethird$, expected
at lower energies, where the radiative lifetimes are longer.  As in the case
of the asymptotic spectral index of $-\onethird$ found for GRB910601rise and
GRB921123, this interpretation would require unexpectedly long radiative 
decay times.  We note, however, that these points are close to the I K-edge,
where deconvolution is particularly difficult.

Most of the GRB show asymptotic low energy spectral indices between 
$-\onethird$ and $\onehalf$.  These indicate intermediate cases between the
index of $-\onethird$ obtained for negligible radiative loss and that of 
$\onehalf$ obtained when the radiating electrons lose all their energy 
during the observation, and are also consistent with adiabatic energy loss.
It is not surprising that the time-integrated spectrum of a complex 
multi-peaked (and probably spatially heterogeneous) event like a GRB should 
lie between these two limiting cases.  Such intermediate behavior is 
consistent with the relativistic shock acceleration model, although it does 
not point to it unambiguously as a value $s = -\onethird$ would.  It is
additional evidence in favor of relativistic shocks that values of $s > 
\onehalf$, although frequently found in other astronomical synchrotron 
sources and readily produced by nonrelativistic shock acceleration processes
(as well as others), have never been found on the low energy side of the 
spectral peak in GRB.
\section{Time Structure}
In a spiky, multi-peaked GRB the synchrotron radiation cooling time (or 
adiabatic expansion time) must be less than the spike duration, and hence 
much less than the burst duration.  Thus spiky bursts are predicted to have 
$s = \onehalf$, unless adiabatic expansion is dominant, in which case $s$
assumes the appropriate value from Equation (3).  Conversely, bursts with 
$s = \onehalf$ must have short radiation times and may have spiky time 
structure.

In a smooth, single peaked GRB the radiation time can be comparable to the 
burst duration because the intensity need not drop to zero again and again 
during the burst.  Hence the predicted instantaneous $s = -\onethird$ may 
approximate the observed spectrum, despite its finite length of integration.

These predictions may be tested by examining the time structures of the GRB
whose spectra we study.  In Figures 2a--d are shown the time histories of 
GRB910601 and GRB921123, which show the best fit to $s = -\onethird$,
GRB930201, which shows the best fit to $s = \onehalf$, and GRB940217, which
resembles $s = \onehalf$, but less closely; GRB940302 is not shown because
of gaps in the data.  In each case the 25--55 KeV count rates are plotted,
as recorded in 64 ms bins; we chose this lowest energy range because only
in it do the spectra approach their low energy asymptotic forms.  The dashed
lines denote the fitted backgrounds and the vertical lines delimit the
intervals over which the spectra shown in Figure 1 were integrated.  
GRB910601rise is the spectrum obtained between the left and middle vertical
lines, and GRB910601 between the left and right vertical lines.
\placefigure{Fig2}

It is evident that the bursts with $s = -\onethird$ are comparatively
smooth, while those with $s = \onehalf$ have complex structure, with many
clearly separated peaks.  In the case of GRB910601 the smooth initial
rising portion of the spectrum has $s = -\onethird$, while the peak of
the burst, intermediate in spikiness between a smooth burst like GRB921123
and very spiky bursts like GRB930201 and GRB940217, has an intermediate
spectrum.  This confirms the predicted qualitative behavior.

It is desirable to quantify the distinction between spiky and smooth bursts.
We modified the algorithm of \cite{LF96} to identify maxima (shown by 
circles) and minima (shown by asterisks) at the ``$7\,\sigma$'' level 
(meaning that minima and maxima must be separated by at least 7 standard 
deviations, so each need be significant only at the $3.5\,\sigma$ level).  
Our modified algorithm searches for broad peaks of comparatively low 
amplitude by smoothing the data between the maxima which it has already 
found, searching again, smoothing further, {\it etc}.  This algorithm is not
perfect, but is useful in demonstrating that the apparent fine structure 
seen, for example, near the maxima of GRB910601 and GRB921123 is probably 
only statistical fluctuation, but that most of the maxima seen by the eye in
GRB930201 and GRB940217 are real.  The number of maxima found is a 
quantitative measure of spikiness.
\section{Discussion}
The spectra we find are all consistent with those predicted by relativistic
shock theory, and we find evidence for its two limiting cases.  In these
limiting cases we confirm the predictions that bursts with $s = -\onethird$ 
should have long radiation times, and therefore smooth profiles, while 
bursts with $s = \onehalf$ should have short radiation times and therefore 
may have spiky multi-peaked profiles.  

Of course, this division of GRB into two classes is not absolute.  It is 
apparent that even the smoother $s = -\onethird$ profiles are not strictly 
single peaked.  This does not contradict the suggestion that the radiation 
time in them is long: even a burst dominated by smoothly varying radiation 
may also contain weaker but more rapidly varying contributions from other 
regions, which do not have a large effect on its spectrum.

The simplest interpretation of our data concerning the correlation between
spectrum and time structure is that the electron radiation time is 
comparable to the width of a spike or subpeak within a burst, or of the
burst itself if it is smooth.  Longer radiation times are excluded by the
observed time dependence (unless adiabatic loss takes the place of energy
loss by radiation).  Shorter radiation times could be consistent with the
observed time structure, if the electrons were continually resupplied at a
steady rate, but this is not consistent with the spectra we observe for
smooth bursts.

The characteristic geometrical time $t_g \sim r/(c\gamma^2)$, where $\gamma$
is the Lorentz factor of bulk motion of the radiating matter and $r$ its
distance from its origin, must in general be comparable to or shorter
than the observed time scale of variation (\cite{SP97b}).  In bursts in
which the instantaneous spectrum ($s = -\onethird$) is observed the time
structure is determined geometrically, and $t_g$ is directly measured.  In
bursts in which the integrated spectrum ($s = \onehalf$) is observed the
time structure only sets an upper bound on $t_g$.

\cite{F95} reported that the characteristic observed subpulse width and
the width of the autocorrelation function of GRB vary as the $\approx - 0.4$
power of the photon energy of observation.  This is in reasonable agreement
with expectations for a population of electrons losing energy by
synchrotron radiation, for which an exponent of $-\onehalf$ is predicted,
and supports our suggestion that subpulse widths are determined by
radiative energy loss.  The same hypothesis also can explain the frequent
observation (\cite{N94}) of fast rise, slow (usually exponential) decay 
time structure in GRB (and the accompanying spectral softening), if the 
rise time is attributed to geometry and the temporal and spectral decay to 
radiative energy loss.  However, it is not possible to explain in this 
manner the few bursts in which the envelope of a rapidly varying intensity
shows slow decay.

The model-dependent procedure (\S2) we use for extracting the source 
spectrum from the observed count distribution is unavoidable, given the
physics of NaI(Tl) detectors.  A skeptic might argue that this invalidates
our conclusions, and we would be unable to prove him wrong.  It is possible,
of course, to use different model spectra in the data reduction, but this
would not solve the problem: if essentially the same results were obtained 
for different spectral models, the skeptic could argue that we had not been 
sufficiently diabolical in our choice of models, while if different results 
were obtained we might argue that that model was unlikely for some reason.
This controversy cannot be resolved without data from detectors of much 
higher intrinsic energy resolution, or over a much broader energy band (from
gamma-rays to visible light, for example).  If these were available then 
deconvolution would not be an issue in determining the broad energy 
distribution of GRB.

Our procedure differs from simply extrapolating the four parameter model 
spectra directly.  The asymptotic form of a spectrum is 
not generally the same as the asymptotic form of a model fitted to the 
spectrum.  The difference may be large when the spectrum does not closely 
approach the asymptote in the region in which the model is fitted to data, 
as in this work.   It is for this reason that we have attempted to determine
the source photon spectrum, rather than just the parameters of the fitted 
model.  For the same reason, the result (\cite{C97}) that the fitted low 
frequency asymptotic slope of the \cite{B93} model varies with time within a
GRB is consistent with our results: the fitted value of this parameter 
depends on the higher frequency spectrum (as is shown by the correlation 
they find between the slope and the spectral break energy), and does not 
uniquely determine the actual low frequency asymptote.

\cite{T96} compared the spectra of five GRB to a theoretical model of the 
instantaneous spectrum which has the same low frequency asymptote as ours, 
but which also describes the higher energy spectrum, and found satisfactory
fits.  His more complex model necessarily has more free parameters; in our 
work the only free parameter is an overall flux normalization.  We tested a 
different hypothesis than his, using data from different detectors on a 
larger sample of GRB; to the extent to which they may be compared, our 
conclusions are consistent with his.

\cite{P96}, using data from the BATSE Spectroscopy Detector, have found in
some GRB a statistically very significant excess flux at photon energies
around 10 KeV, lower than the lowest energy for which we have data.  These 
excess fluxes are not consistent with our power law spectra (as shown by 
their fitted parameters), or with any simple extrapolation from higher 
energies.  An additional radiation mechanism may be required, as discussed
by \cite{P96}; this problem is beyond the scope of this paper.

Relativistic shocks are encountered elsewhere in astrophysics, for example,
in the termination shocks of pulsar winds and relativistic jets.  It is
unclear whether the spectra of these objects have the predicted slopes.  If 
their radiation is emitted by particles accelerated in their central sources
then their spectra need not resemble those we have discussed; these 
particles flow out into the wind and radiate there (their
adiabatic expansion ``losses'' are only the conversion of an isotropic
distribution of velocities to directed motion---collimation---preserving
their mean energy and spectrum).  If, however, the radiation is emitted by 
particles accelerated in relativistic termination shocks, as is plausible
in the lobes of double radio sources and in supernova remnants which contain
pulsars, then the characteristic spectrum of relativistic shock acceleration
is expected.  If these sources are observed in steady state, as would be 
expected for a relativistic gas with nonrelativistic bulk velocity, with 
ages much greater than the synchrotron lifetimes of the radiating electrons,
then a spectral index $s = \onehalf$ is predicted.  This may account for the
frequent observation of spectral indices close to this value.

A spectral index of $\onehalf$ is obtained for any energy distribution of
accelerated electrons, if they are observed throughout the radiative decay
of their energy, so long as their initial characteristic synchrotron
frequencies exceed the range over which the spectral index is measured.
This value of $s$ is therefore expected even when the accelerating mechanism
is not a relativistic shock.  It is only necessary that the condition 
discussed by \cite{K94a} be met---that the distribution of accelerated
electron energies be flat enough ($dN/dE \propto E^{-p}$ with $p < 
\onethird$).  Observed spectral indices $\approx 0.5$ therefore need not 
imply acceleration by nonrelativistic shocks with $p \approx 2$.

The spectral index $s$ affects the extrapolation of GRB spectra to lower
frequencies, and particularly their predicted visible magnitudes.  For
example, replacement of $s = -\onethird$ by $s = \onehalf$ in extrapolating
from 300 KeV gamma-rays to visible light would increase the visible 
brightness by about 10 magnitudes.  For the $10^{-5}$ erg/cm$^2$s ``burst of
the month'' discussed by \cite{K94a} the predicted visible magnitude would 
then be about 8, rather than 18.  Similarly, \cite{FB96} extrapolated
the fitted four-parameter model either all the way to visible frequencies
or to 1 KeV, below which an $s = -\onethird$ extrapolation was assumed, and 
found a very broad range of predicted visible magnitudes, including some 
comparatively bright values.  

Magnitudes much less than 18 are unduly optimistic.  The synchrotron
lifetime increases as the electron energy declines.  Therefore, even if the 
gamma-ray region of the spectrum had $s = \onehalf$, the spectral index 
would be predicted to approach the instantaneous value $s = 
-\onethird$ at lower frequencies, probably long before the visible region.
Similarly, if the spectral index in the gamma-ray region reflects adiabatic 
expansion ($-\onequarter \le s \le \onequarter$), as the expansion proceeds 
and the electron energies decline the expansion time scale becomes longer 
and the spectral index will again approach its instantaneous value $s = 
-\onethird$.  In either case, the consequence is that the predicted visible 
magnitude is less than 18, but probably by a few magnitudes rather than by 
10.  The same conclusion can be reached without any theoretical assumptions 
at all by noting that the spectrum has not approached its asymptotic form at
$h\nu = 300$ KeV, but may at X-ray energies.
\acknowledgments
We thank B.~E.~Schaefer and G.~N.~Pendleton for very useful discussions.
J.~I.~K.~thanks Washington University for the grant of sabbatical leave,
the Hebrew University for hospitality and a Forchheimer Fellowship.  This
research was partially supported by NASA NAG 3516, NAG 52682 and NAS
8-36081, NSF AST 94-16904 and an Israel-US BSF grant, and has made use of 
data obtained through the Compton Gamma Ray Observatory Science Support 
Center Online Service, provided by the NASA Goddard Space Flight Center.

\newpage
\figcaption{Spectra of 11 GRB.  Solid lines correspond to $s = -\onethird$
and dashed lines to $s = \onehalf$.  GRB910601rise denotes the spectrum of
the initial (pre-maximum) rise of GRB910601; see Figure 2. \label{Fig1}}
\figcaption{Intensity histories of four GRB in the 25--55 KeV band.  The 
dashed lines are fits to the backgrounds.  Circles denote maxima and 
asterisks minima.  The spectra in Figure 1 were integrated over the periods
between the vertical lines.  The spectrum labeled GRB910601rise was
integrated from the first to the middle line, and that labeled GRB910601
from the first to the last line. \label{Fig2}}
\begin{figure}
\epsscale{0.5}
\plottwo{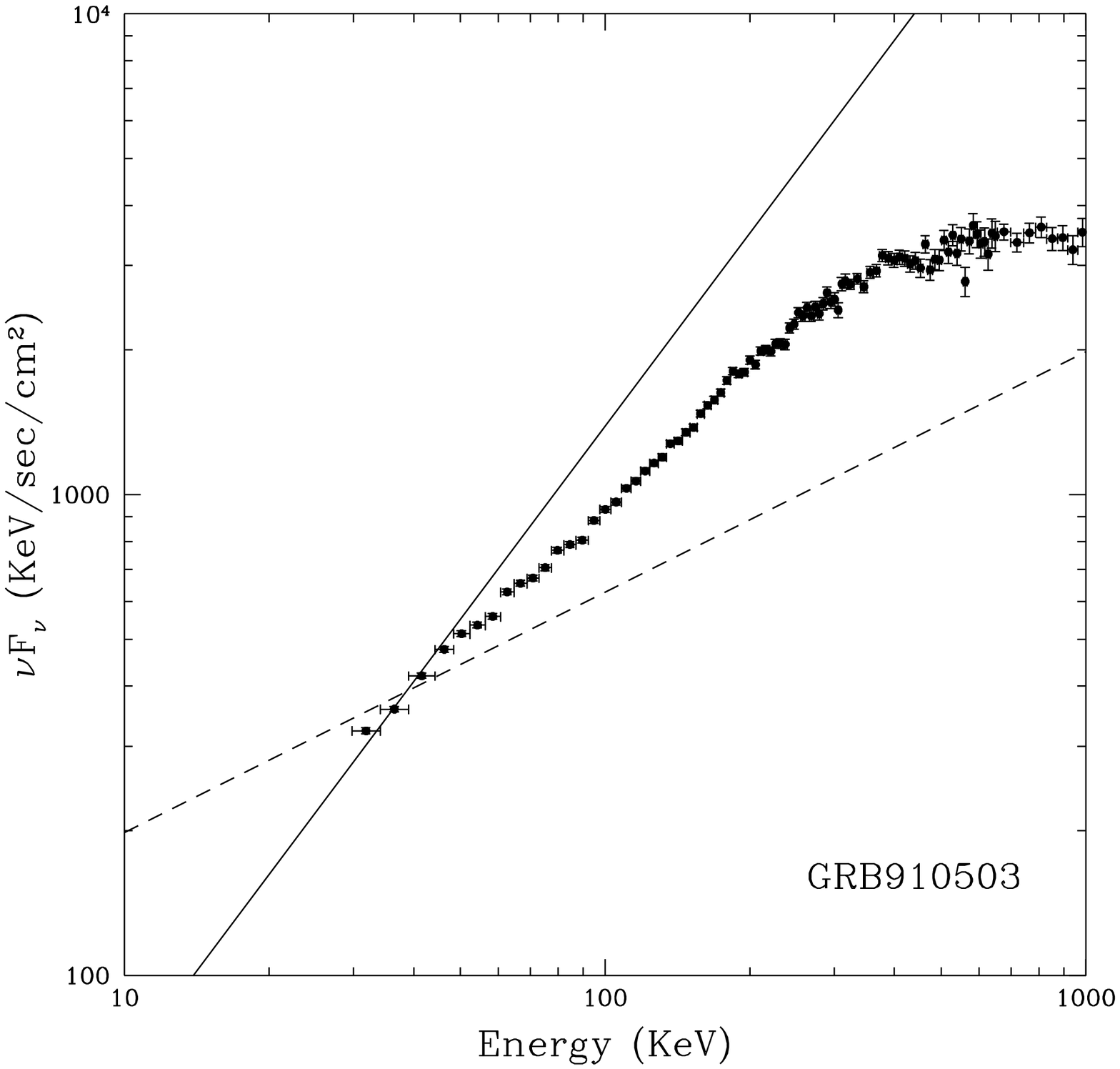}{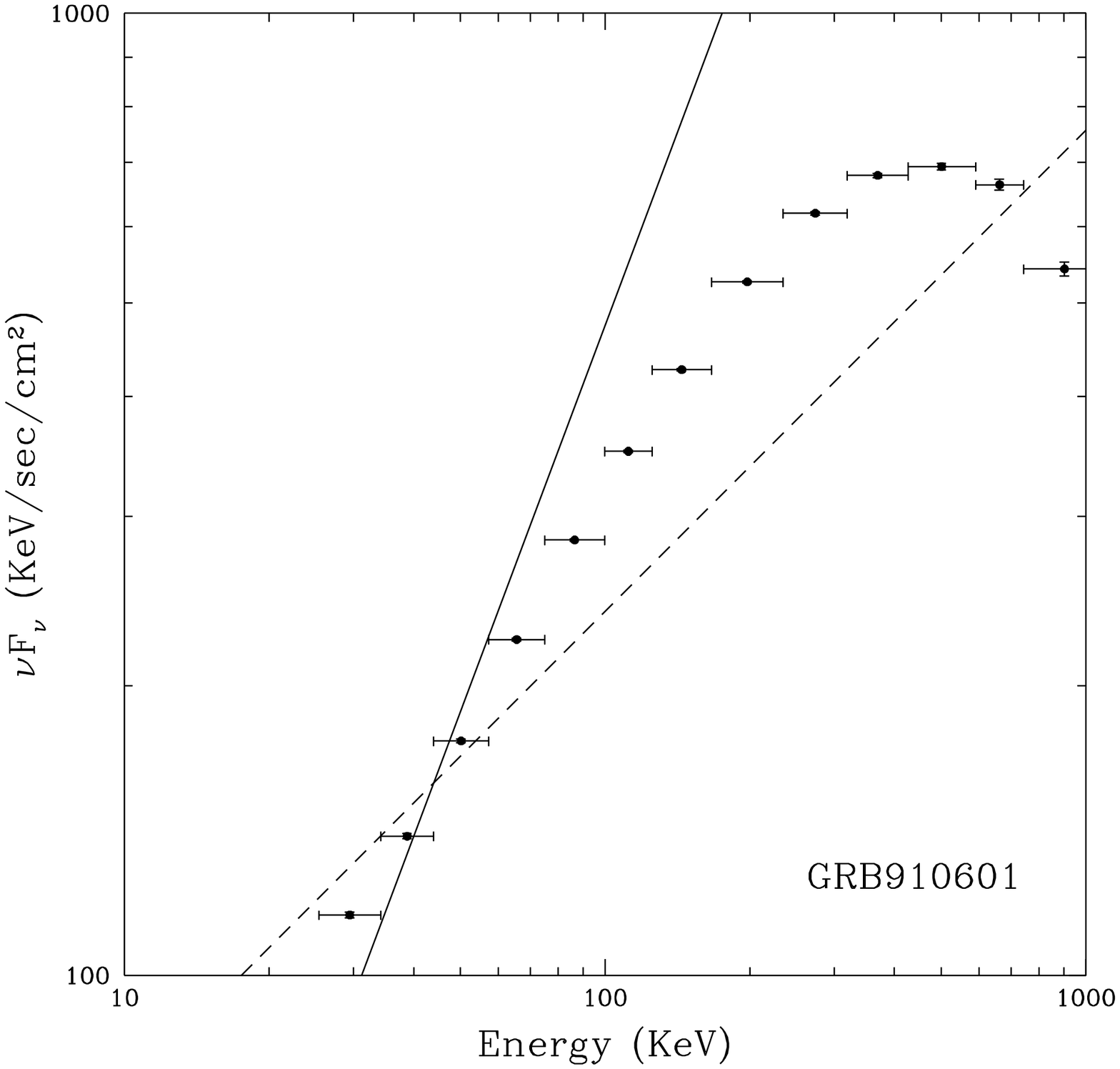}
\plottwo{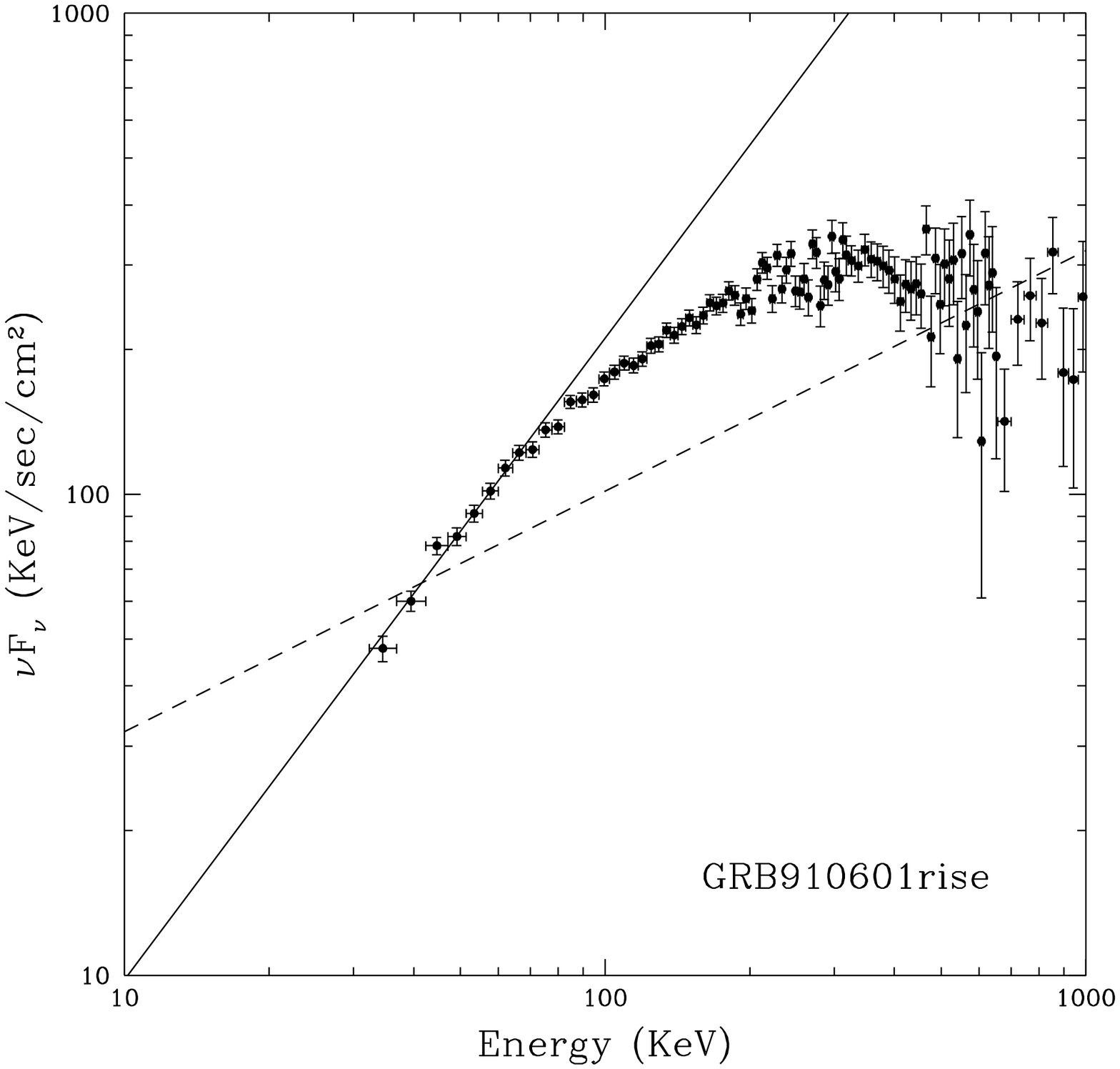}{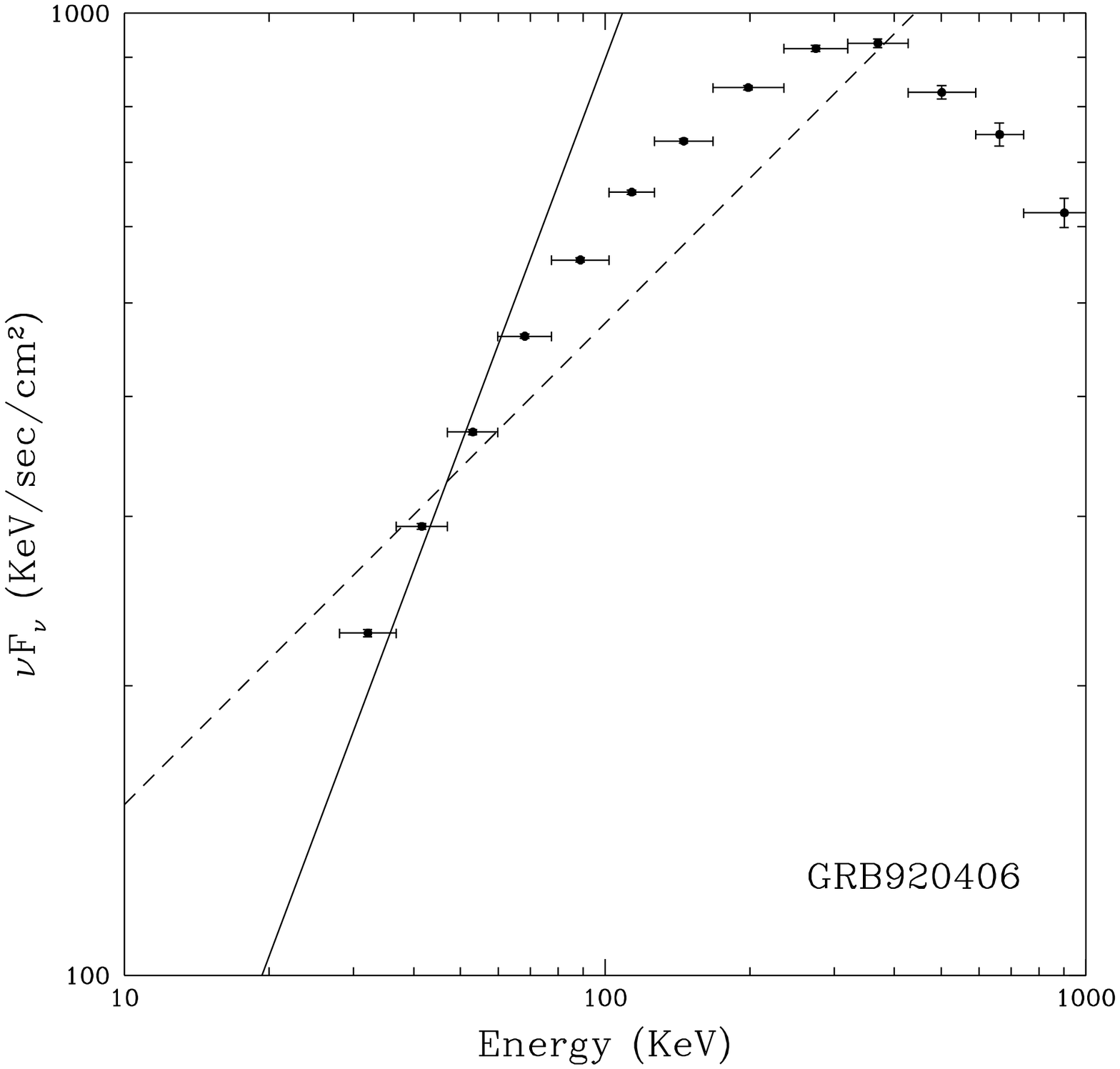}
\end{figure}
\newpage
\begin{figure}
\epsscale{0.5}
\plottwo{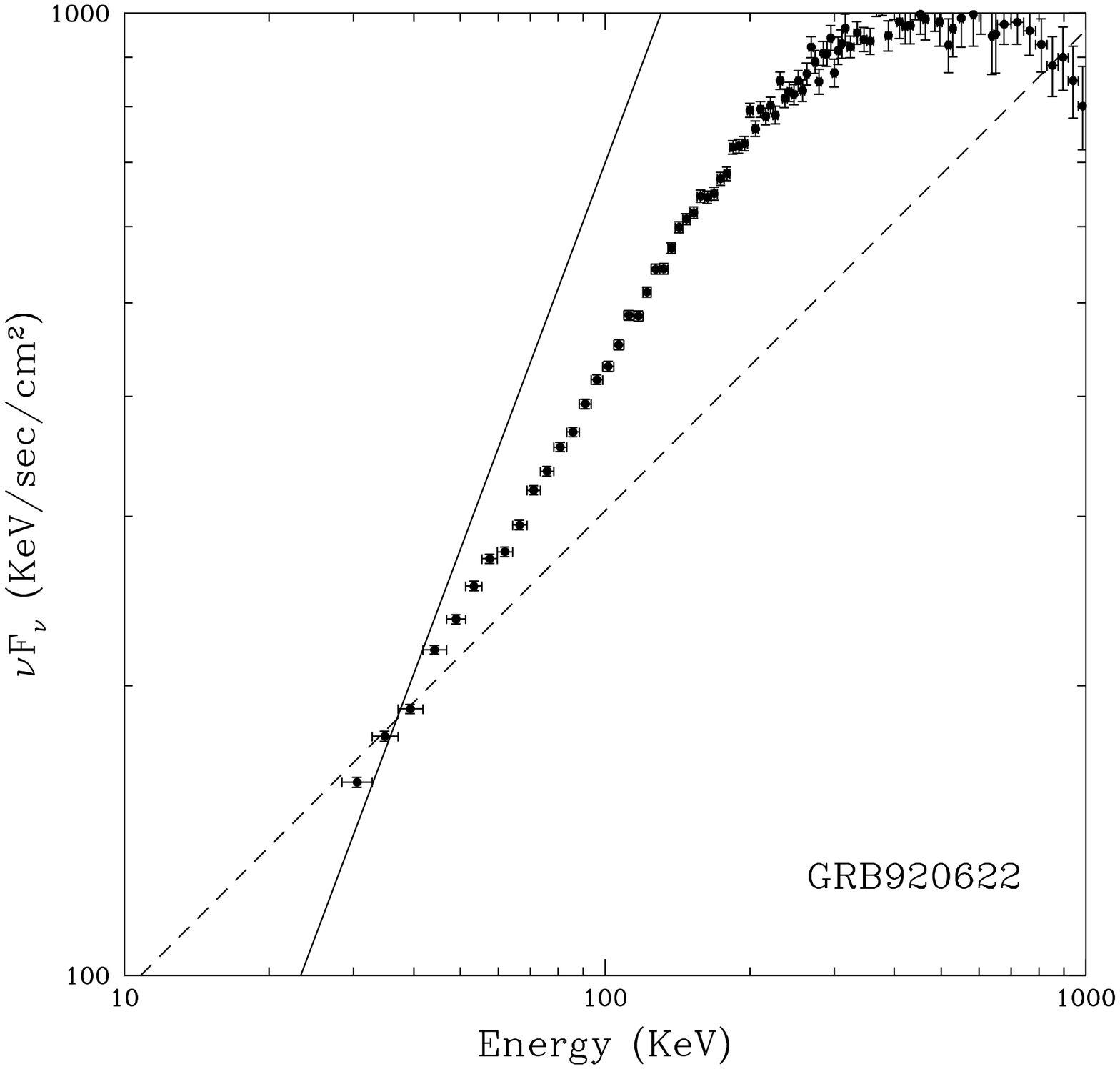}{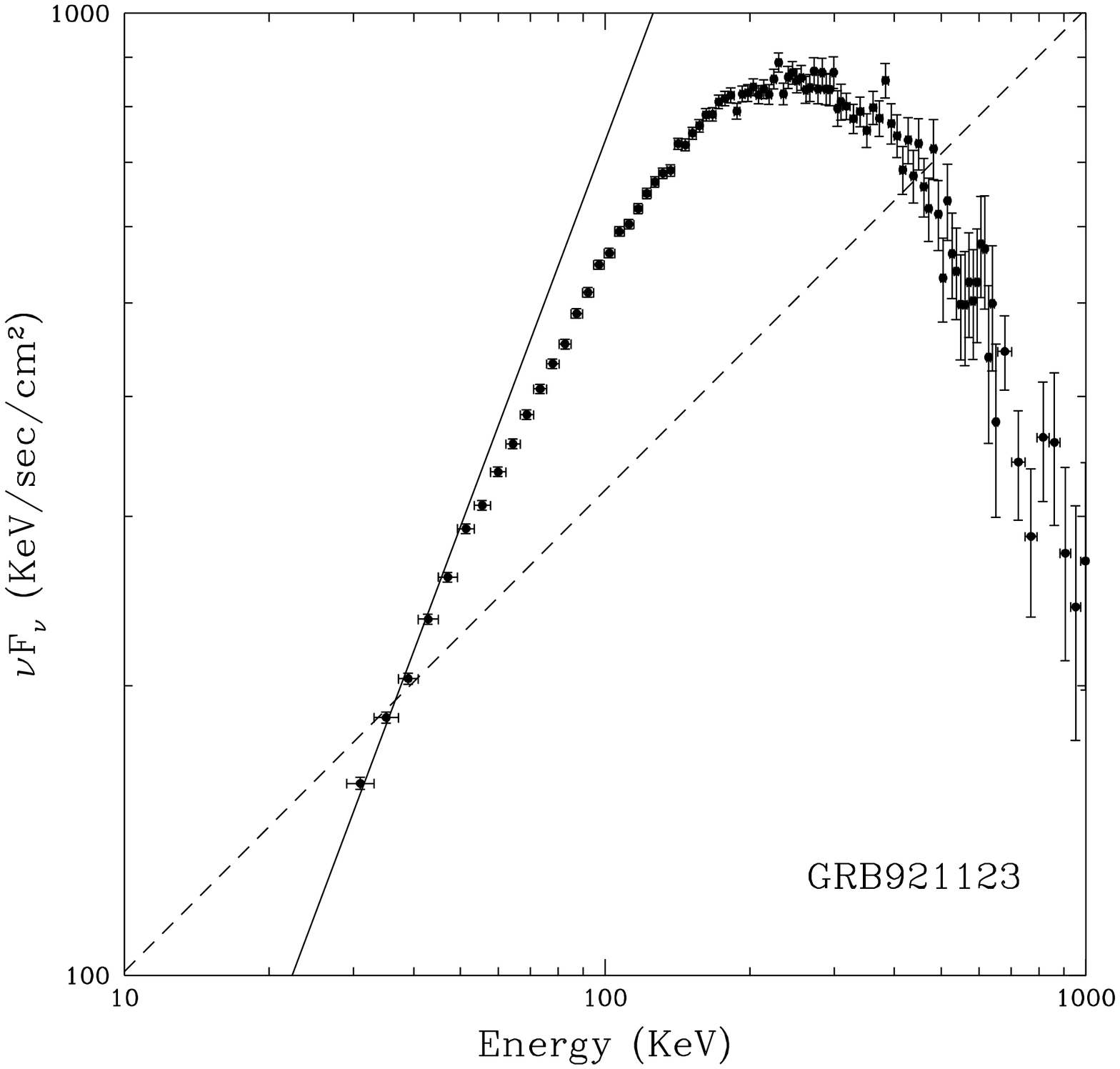}
\plottwo{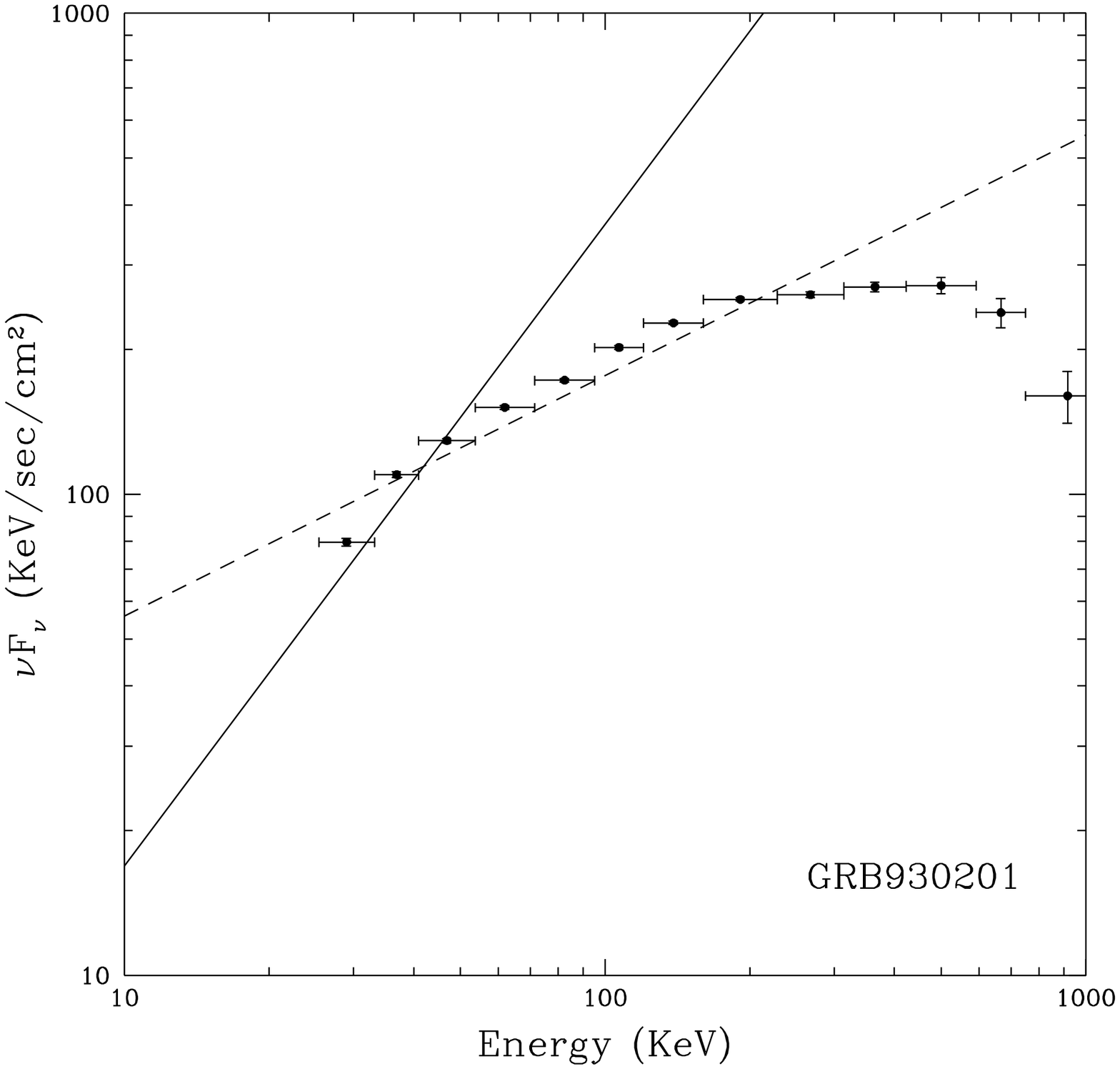}{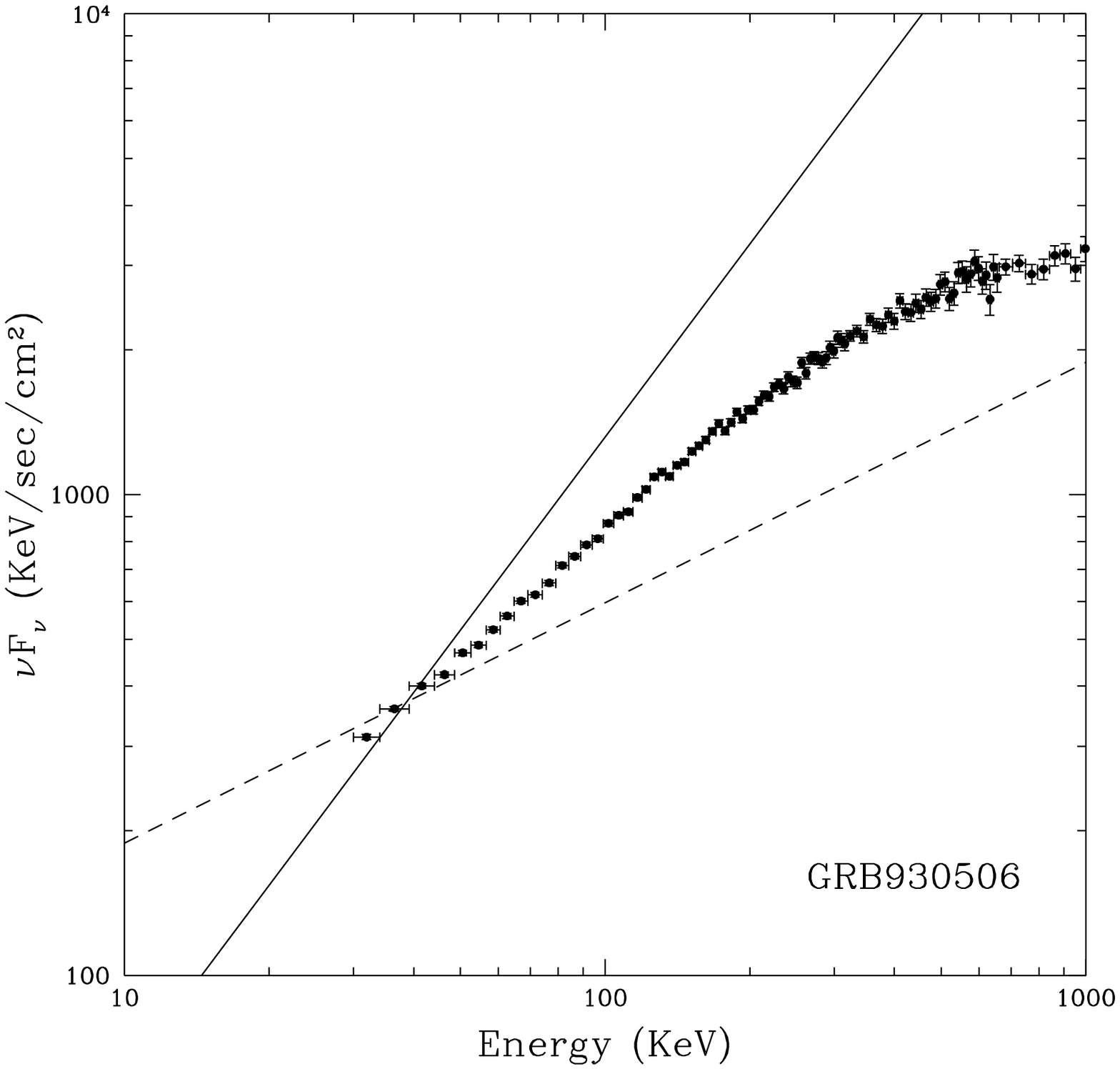}
\end{figure}
\newpage
\begin{figure}
\epsscale{0.5}
\plottwo{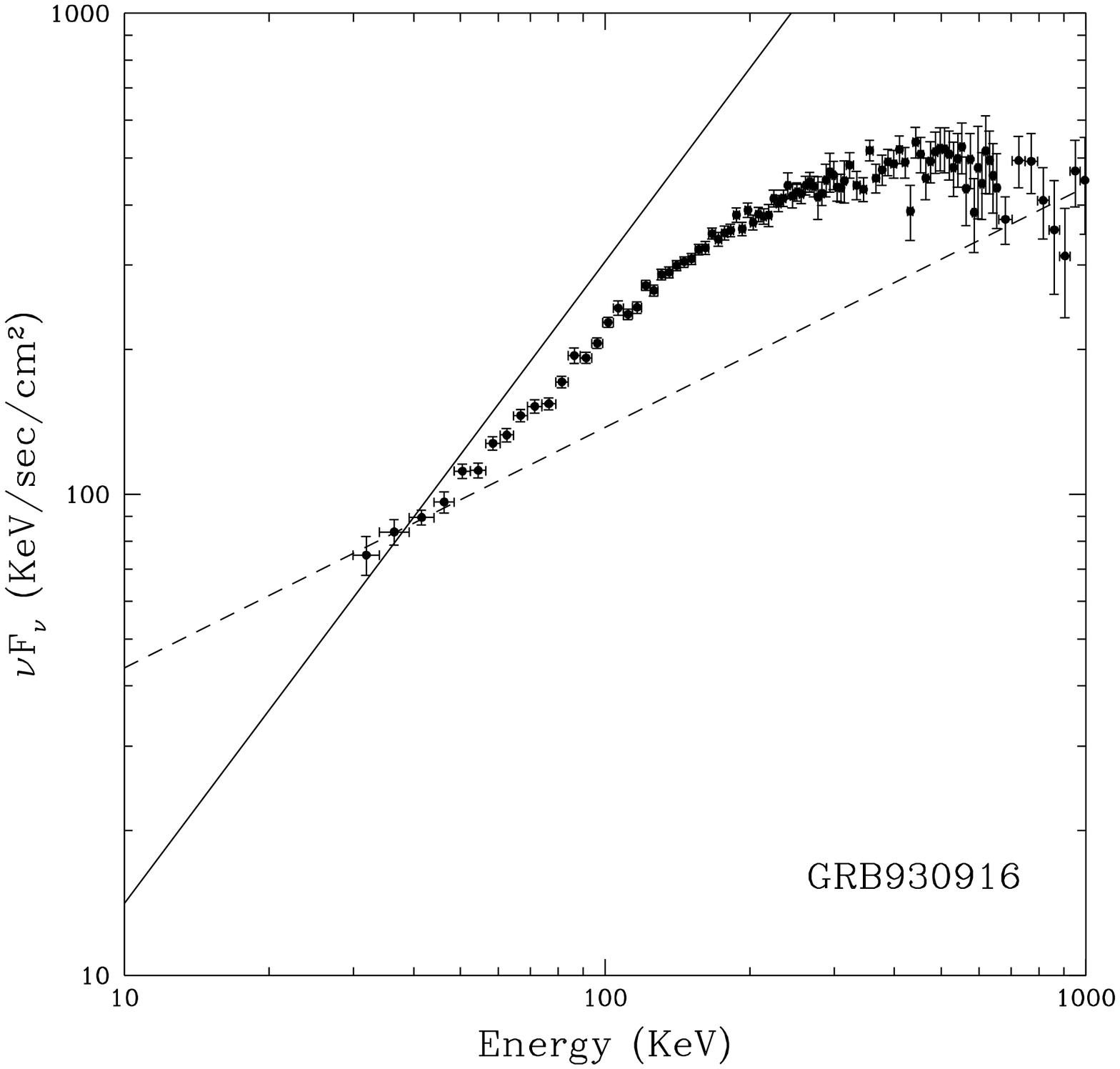}{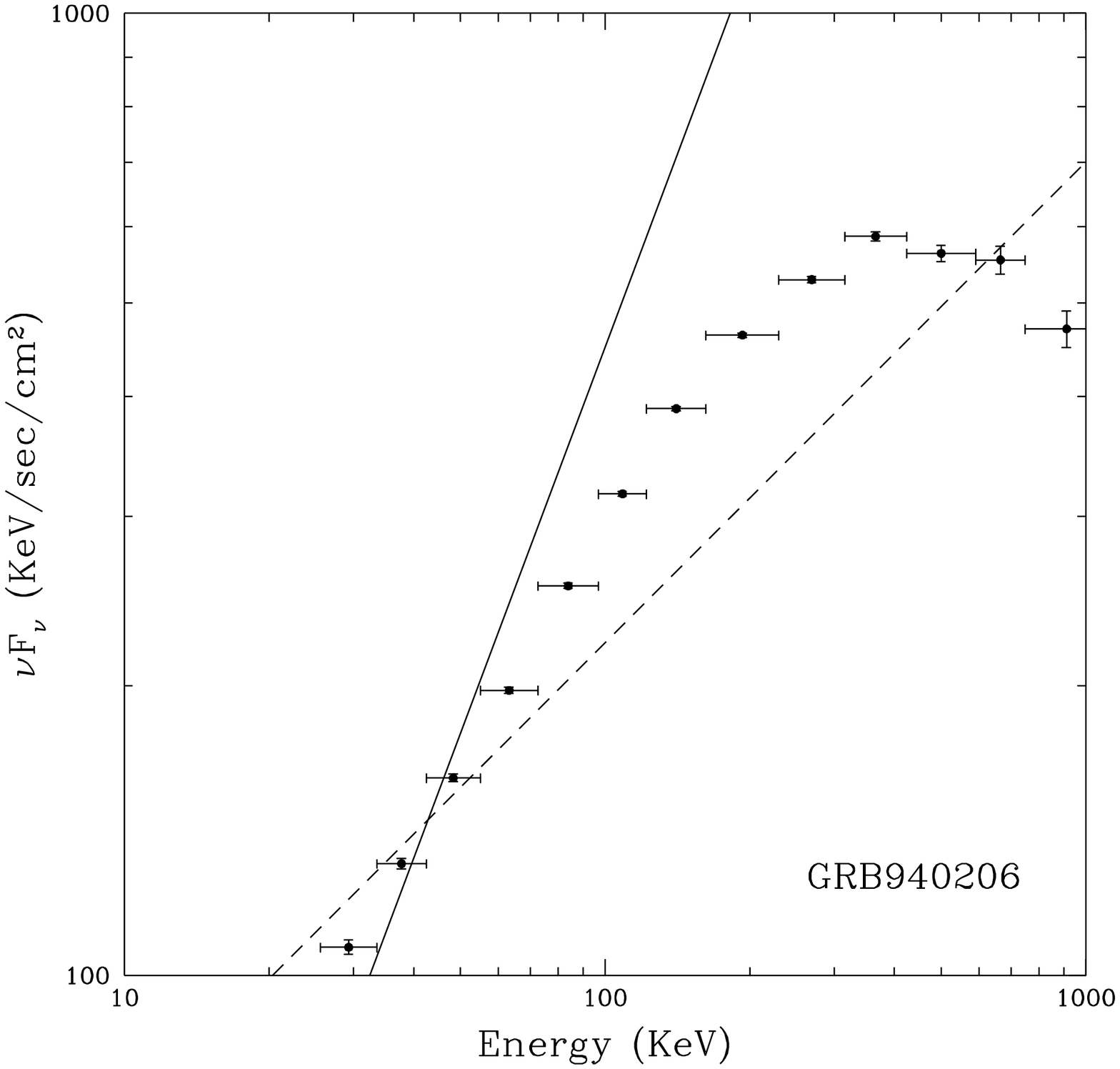}
\plottwo{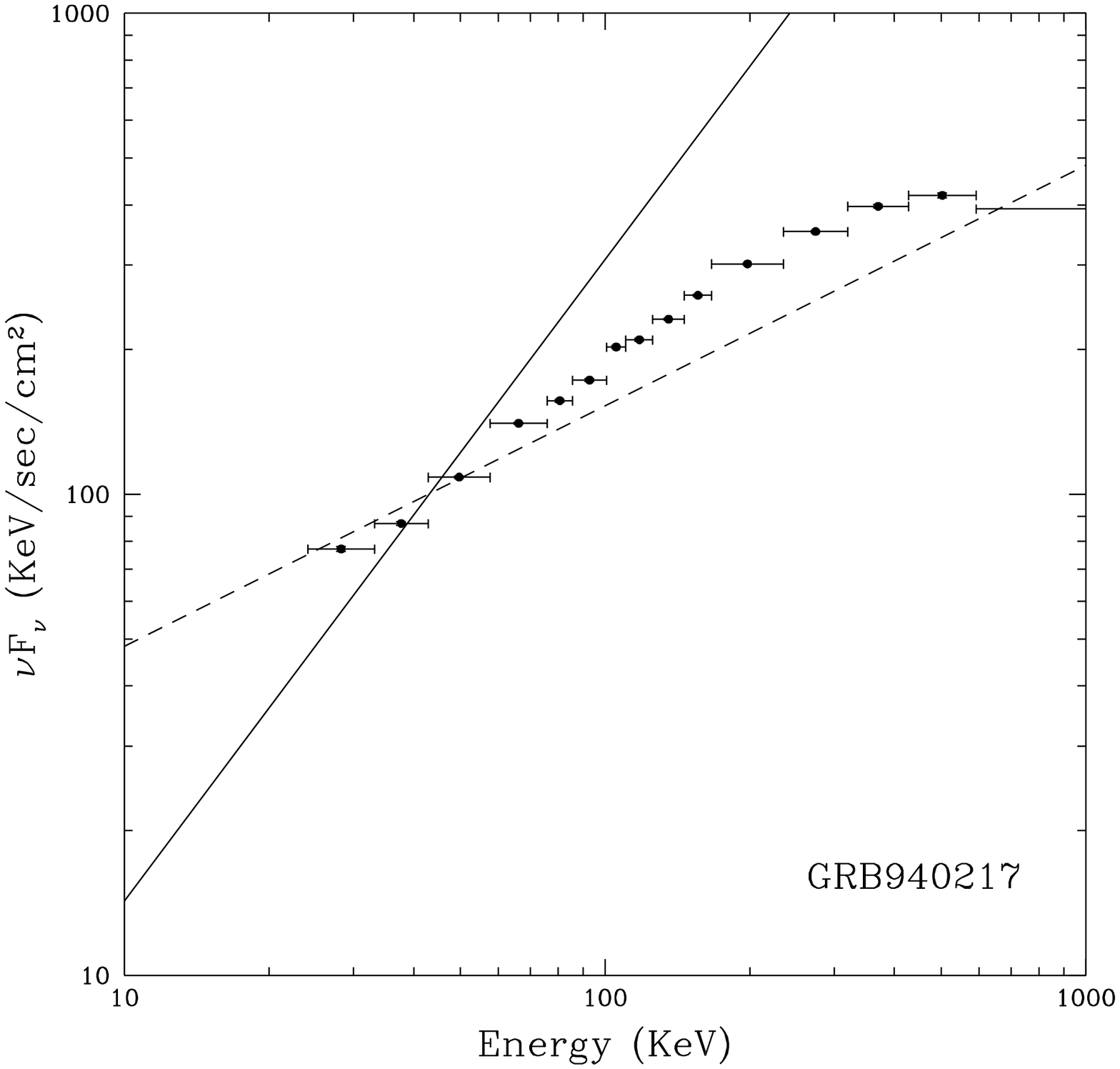}{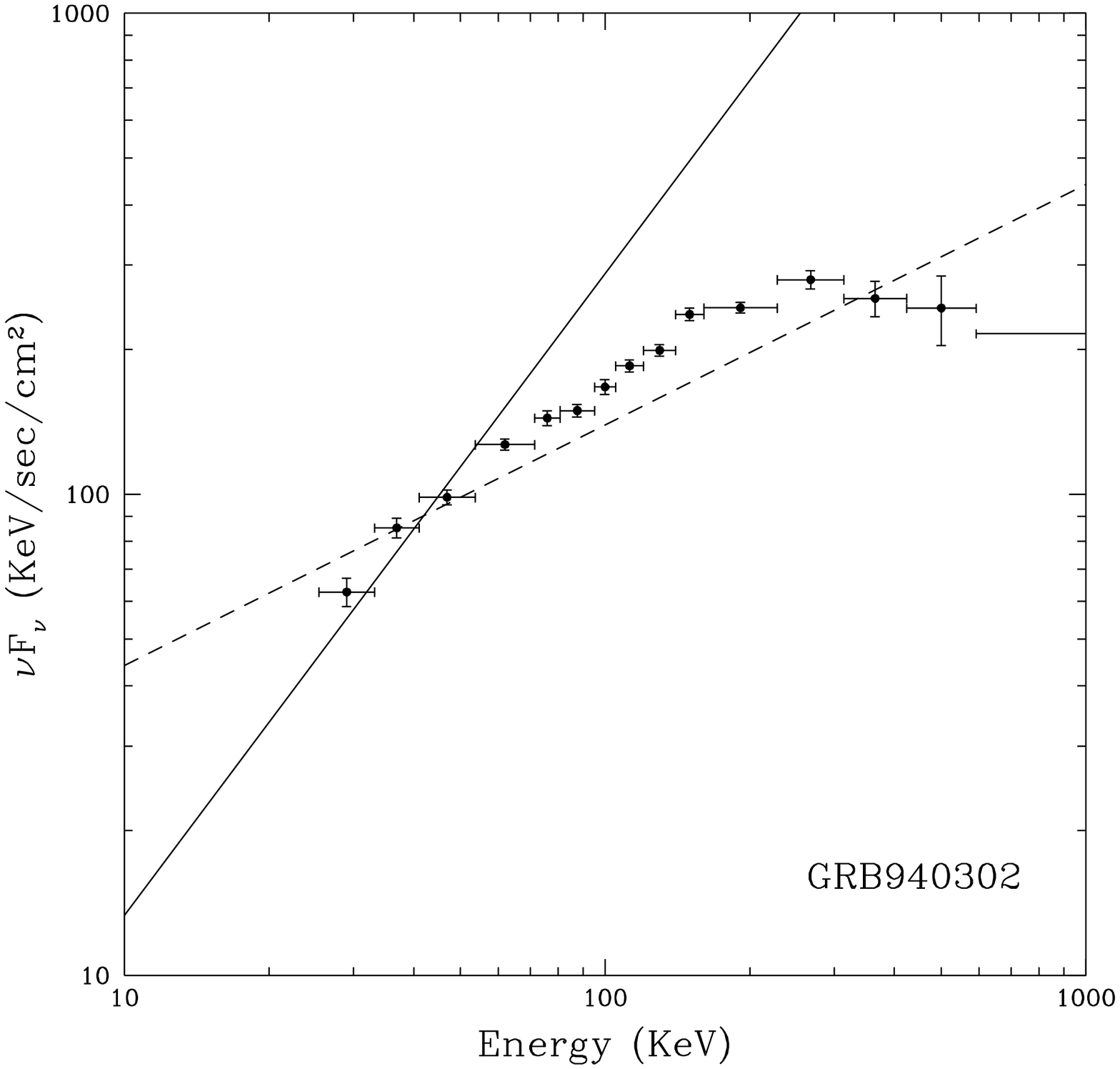}
\end{figure}
\newpage
\begin{figure}
\epsscale{0.5}
\plottwo{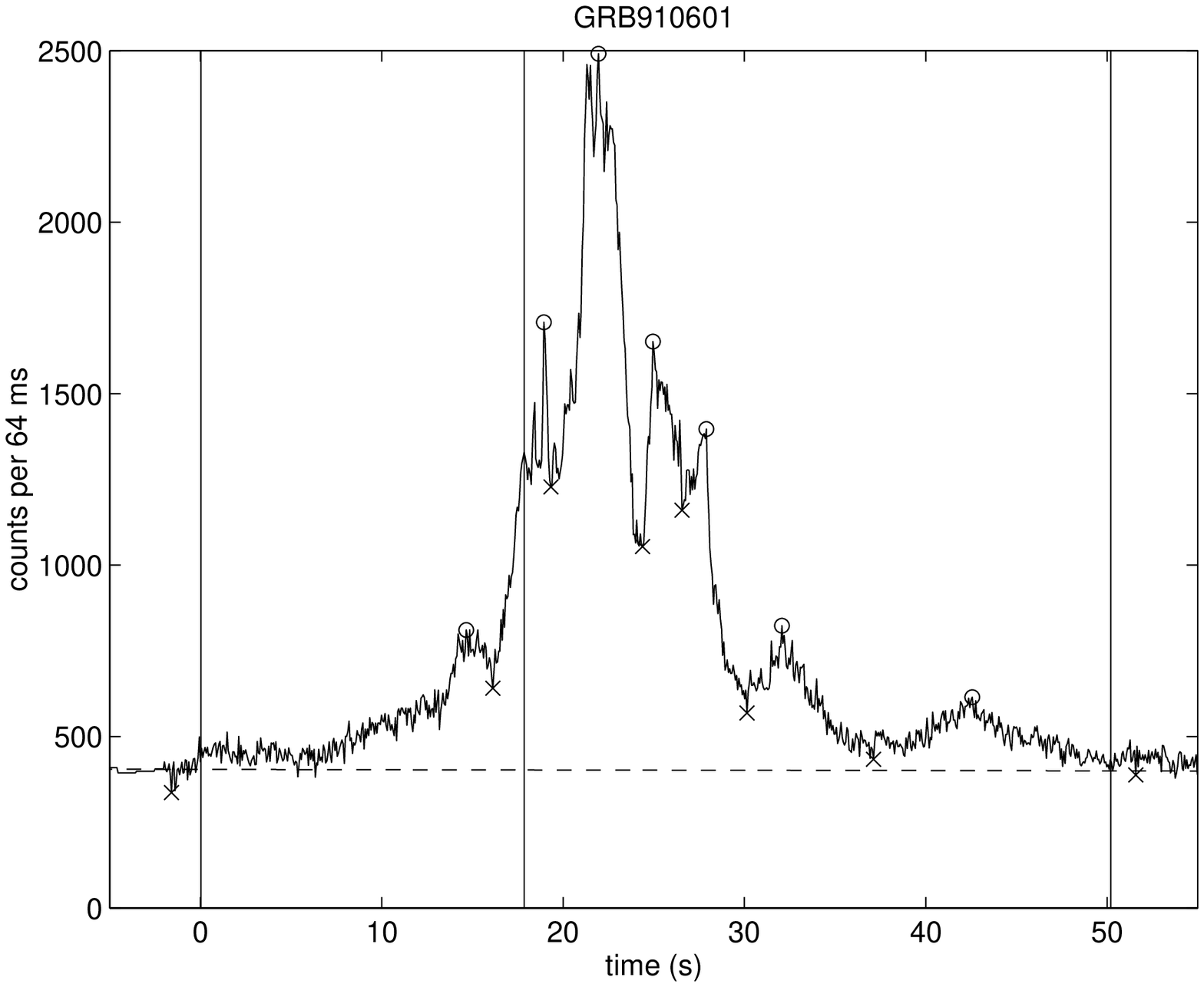}{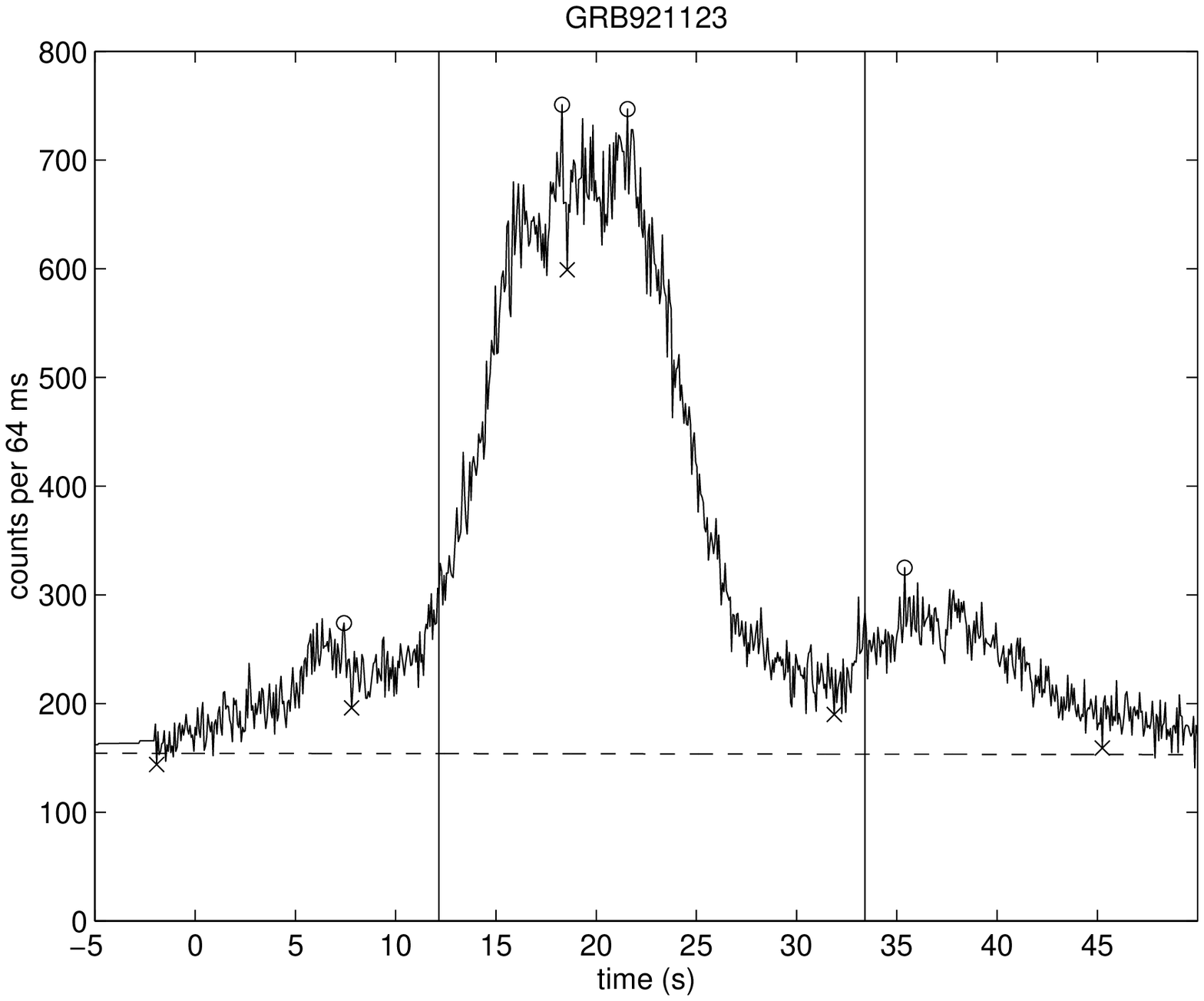}
\plottwo{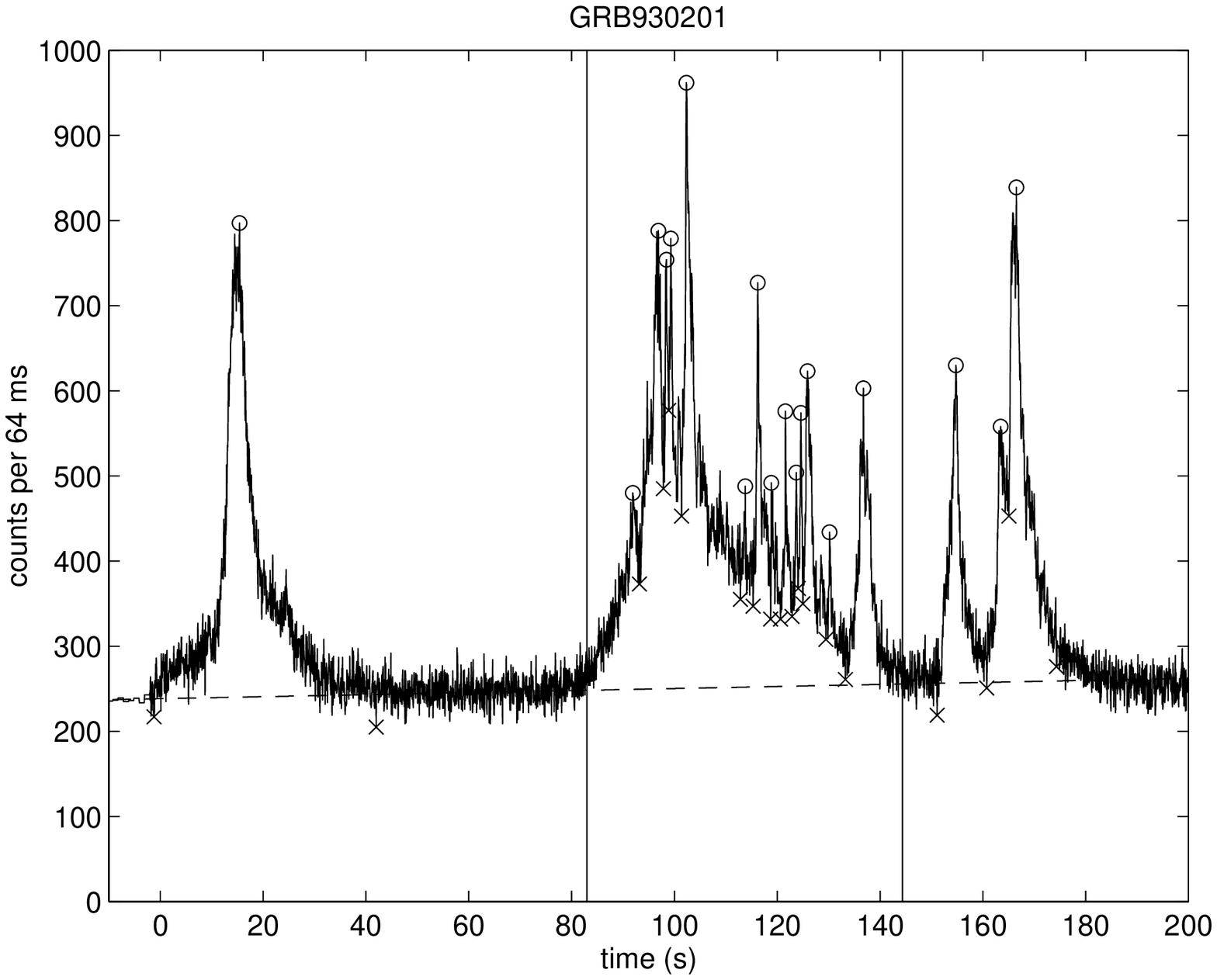}{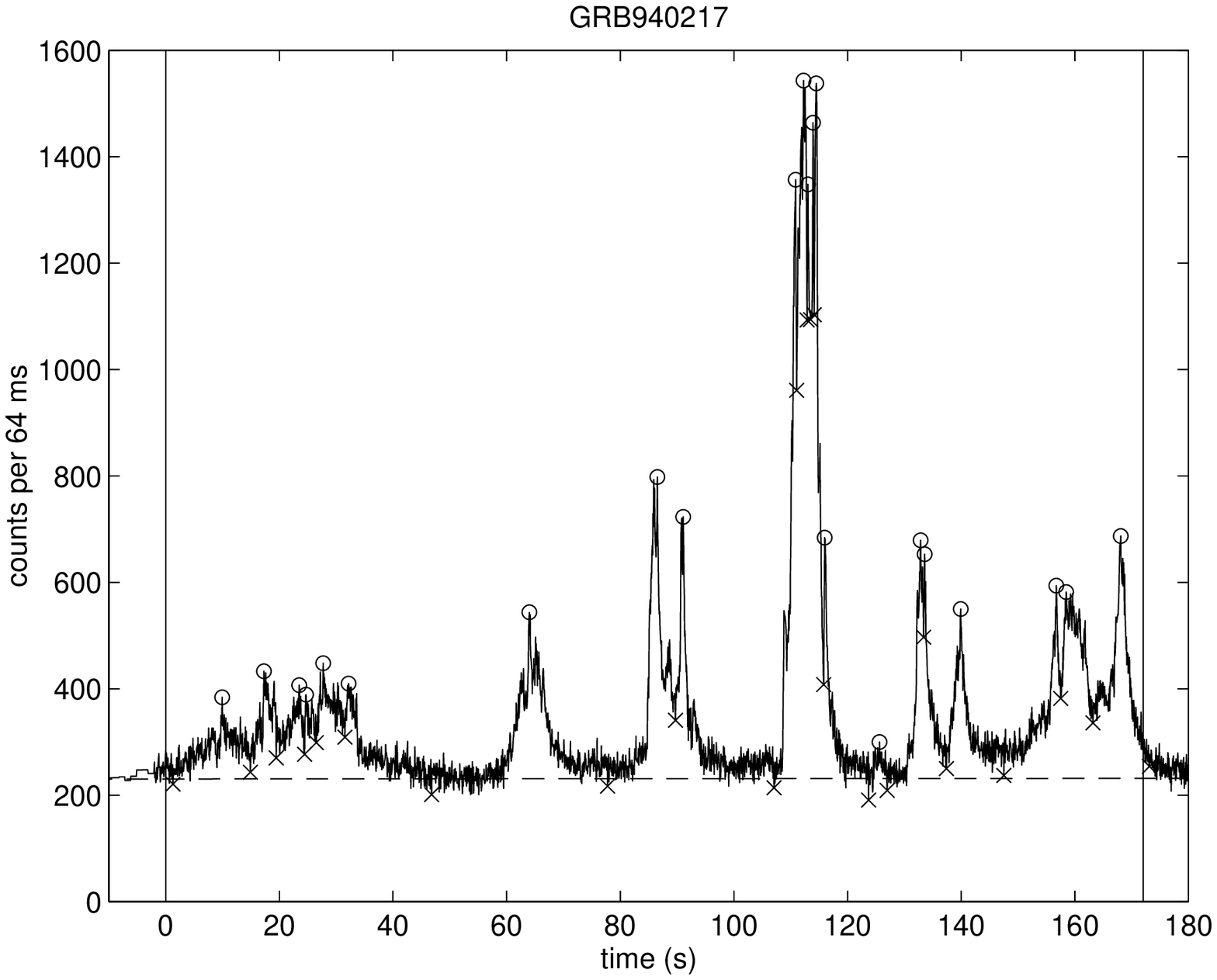}
\end{figure}

\begin{thebibliography}{}
\bibitem[Band, {\it et al.} 1993]{B93} Band, D., {\it et al.} 1993, \apj\
413, 281
\bibitem[Briggs 1996]{B96} Briggs, M. S. 1996, Gamma-Ray Bursts, 3rd
Huntsville Symposium, eds. C. Kouveliotou, M. S. Briggs and G. J. Fishman
(New York: AIP), p. 133
\bibitem[Crider, {\it et al.} 1997]{C97} Crider, A. {\it et al.} 1997, \apj\
in press
\bibitem[Fenimore, {\it et al.} 1995]{F95} Fenimore, E. E., in't Zand, J. J.
M., Norris, J. P., Bonnell, J. T. \& Nemiroff, R. J. 1995, \apjl\ 448, L101
\bibitem[Ford \& Band 1996]{FB96} Ford, L. A. \& Band, D. L. 1996, \apj\ 
473, 1013
\bibitem[Horack 1991]{H91} Horack, J. M. 1991, Development of the Burst and
Transient Source Experiment (BATSE), NASA Ref. Pub. 1268
\bibitem[Katz 1994a]{K94a} Katz, J. I. 1994a, \apjl\ 432, L107
\bibitem[Katz 1994b]{K94b} Katz, J. I. 1994b, \apj\ 422, 248
\bibitem[Li \& Fenimore 1996]{LF96} Li, H. \& Fenimore, E. E. 1996, \apjl\
469, L115
\bibitem[Meegan, {\it et al.} 1996]{M96} Meegan, C. A., {\it et al.} 1996,
\apjs\ 106, 65
\bibitem[Nemiroff, {\it et al.} 1994]{N94} Nemiroff, R. J., Norris, J. P.,
Kouveliotou, C., Fishman, G. J., Meegan, C. A. \& Paciesas, W. S. 1994,
\apj\ 423, 432
\bibitem[O'Kelley 1961]{OK61} O'Kelley, G.~D.~1961, Methods of Experimental
Physics 5A, eds.~L.~C.~L.~Yuan \& C.--S.~Wu (New York: Academic Press), 616
\bibitem[Pendleton, {\it et al.} 1995]{P95} Pendleton, G. N., {\it et al.}
1995, Nucl. Instr. and Meth. in Phys. Res. A364, 567
\bibitem[Preece, {\it et al.} 1996]{P96} Preece, R. D., {\it et al.} 1996,
\apj\ 473, 310
\bibitem[Preece, {\it et al.} 1997]{P97} Preece, R. D., {\it et al.} 1997,
in preparation
\bibitem[Sari, Narayan \& Piran 1996]{SNP96} Sari, R., Narayan, R. \& Piran,
T. 1996 \apj\ 473, 204
\bibitem[Sari \& Piran 1997a]{SP97a} Sari, R. \& Piran, T. 1997a, \mnras\ in
press
\bibitem[Sari \& Piran 1997b]{SP97b} Sari, R. \& Piran, T. 1997b, \apj\ in
press (astro-ph/9701002)
\bibitem[Schaefer, {\it et al.} 1994]{S94} Schaefer, B. E., {\it et al.} 
1994, \apjs\ 92, 285
\bibitem[Tavani 1995]{T95} Tavani, M. 1995, \apss\ 231, 181
\bibitem[Tavani 1996ab]{T96} Tavani, M. 1996a, \prl\ 76, 3478
\bibitem[]{T96b}Tavani, M. 1996b, \apj\ 466, 768
\end{thebibliography}
\end{document}